\documentclass[12pt]{article}
\usepackage{graphicx}
\usepackage{psfig}
\usepackage{epsfig}
\usepackage{epsf}
\usepackage{amssymb}
\usepackage{rotate}

\def\th13 {\theta_{13}}

\def\en {E_{\nu}}

\def\ens {{E_{\nu}^\ast}}

\def\lp {\lambda'}
\newcommand{\beq}{\begin{equation}}
\newcommand{\eeq}{\end{equation}}
\def\bea{\begin{eqnarray}}
\def\eea{\end{eqnarray}}
\def\benu{\begin{enumerate}}
\def\eenu{\end{enumerate}}
\newcommand{\no}{\nonumber}
\begin{document}
\thispagestyle{empty}
\begin{flushright}
\texttt{hep-ph/0609252}\\
\texttt{HRI-P-06-09-001}\\
\texttt{DO-TH~06/09}\\
\texttt{CU-PHYSICS-15/2006}\\
\end{flushright}
\vskip 30pt
\parindent 0pt

\begin{center}
{\Large {\bf Probing Lepton Number Violating Interactions with
Beta-beams}}

\vskip 20pt
\renewcommand{\thefootnote}{\alph{footnote}}

\large{\bf{ Sanjib Kumar Agarwalla$\footnote{ E-mail address:
sanjib@mri.ernet.in}^{\dagger}$, Subhendu Rakshit$\footnote{E-mail
address: rakshit@zylon.physik.uni-dortmund.de}
^{\ddagger}$}}\\
\large{\bf{ and Amitava Raychaudhuri$^{\dagger}$}}
\end{center}
\begin{center}
\small$\phantom{i}^{\dagger}${\em Harish-Chandra Research Institute,\\
Chhatnag Road, Jhunsi, Allahabad - 211~019, India \\ and \\
Department of Physics, University of Calcutta, \\
Kolkata - 700~009, India}\\~\\
\small$\phantom{i}^{\ddagger}${\em Universit\"{a}t Dortmund, Institut f\"{u}r Physik,}\\
{\em D-44221 Dortmund, Germany} \\
\vskip 30pt

%%%%%%%%%%%%%%%%%%%%%%%%%%%%%%%%%%%%%%%%%%%%%%%%%%%%%%%%%%%%%%%%%%%%%%%%%%%%%%%%%%%%%%%%%%%
{\bf ABSTRACT}

\vskip1.0cm

\end{center}
We show that a detector placed near a beta-beam storage ring can probe
lepton number violating interactions, as predicted by supersymmetric
theories with $R$-parity non-conservation. In the presence of
such interactions, $\nu_\tau$ can be produced during
$\beta$-decay leading to tau leptons  through weak interactions.
Alternatively, electron neutrinos from $\beta$-decay of
radioactive ions can produce tau leptons in a nearby detector
through these interactions. The muons from the decay of these
tau leptons can be readily identified in a small iron calorimeter
detector and will signal violation of $R$-parity.

\newpage

\renewcommand{\thesection}{\Roman{section}}
\setcounter{footnote}{0}
\renewcommand{\thefootnote}{\arabic{footnote}}
%\renewcommand{\thefootnote}{\fnsymbol{footnote}}

%%%%%%%%%%%%%%%%%%%%%%%%%%%%%%%%%%%%%%%%%%%%%%%%%%%%%%%%%%%%%%%%%%%%%%%%
\section{Introduction}\label{sec1}

In the standard model (SM), lepton number ($L$) conservation is only
accidental; the particle content and the requirement of
renormalizability ensure that each lepton flavour number is conserved
separately. However, non-zero neutrino masses, as indicated by recent
neutrino oscillation experiments, have proved that the success of the
SM should be viewed as that of a low energy effective theory. It is
not unreasonable to expect that in some extensions of the SM, $L$
conservation may not hold. Indeed, a Majorana mass term for the
neutrinos violates total lepton number.  The non-observation of direct
$L$ violation in the past experiments have put stringent constraints
on some of these interactions. In this letter we show that
beta-beams~\cite{zucc,mezzetto,burguet,agarwalla1,agarwalla2,review,Volpeadd}
and a nearby detector can be a good further probe of such
interactions.\\

A beta-beam consists of a high intensity collimated beam of electron
neutrinos produced from the beta decay of boosted radioactive
ions. The recent progress in nuclear physics experimental techniques
allows the design of beta-beams of high luminosity so that it can have
comparable physics potential as that of proposed super-beam upgrades
or even neutrino factories, which are technologically challenging at
present. These beams may have widespread applications in particle
physics, nuclear physics and astrophysics. The high collimation
achievable with these beams allows neutrino oscillation experiments
with long baselines\footnote{In~\cite{agarwalla2} it was pointed out
that these beams while traversing a long base-length can get
influenced by $L$-violating interactions through matter effects, thus 
polluting oscillation signals.}. However, for other physics
studies, a small detector placed close to the source has been
proposed. For this work, the advantage of a `near' detector is
twofold. Firstly, due to the short base-length, neutrinos do not get
much scope to oscillate before being detected, which could otherwise
mimic signals of the $L$-violating interactions ($\not{\! L}$).  The
other obvious advantage is that a larger part of the beam can be
picked up with a smaller detector.\\

We consider placing a 5 kT cylindrical detector, aligned with the
beam axis, within 1 km from the beta-beam storage ring. The
$L$-violating interactions can lead to tau leptons in
near-detector experiments in two ways. A $\nu_\tau$ can be
produced due to such interactions during $\beta$-decay, yielding
a $\tau$ through weak charged current interactions in the
detector. Alternatively, the electron neutrinos in the beam,
produced through usual $\beta$-decay, can undergo $L$-violating
interactions with the detector, leading to tau leptons. The taus 
promptly decay, part of the time in a muonic channel. Iron
calorimeters with active detector elements serve well for
identifying these muons, which leave long tracks in the detector,
and for filtering out backgrounds.  We will also briefly comment
on water \u{C}erenkov and other detectors.\\

In the following section we present a brief account of the
experimental setup. In section~\ref{sec3}, $\not{\! L}$ interactions
are discussed in the context of $R$-parity violating (RPV)
supersymmetry~\cite{couplings,latest_bounds}. We stress how
$\beta$-decay can be affected in the presence of such interactions,
yielding $\nu_\tau$ in a few cases in place of the standard $\nu_e$.
We also describe the processes via which $\nu_e$ produce tau leptons
in the detector. The expected number of muon events from tau decay and
the constraints ensuing in the event of their non-observation will be
presented in section~\ref{sec4}.

\section{Beta-beam flux at a near-detector}\label{sec2}

The proposal of a beta-beam was put forward by
Zucchelli~\cite{zucc}. It is based on the concept of creating a
pure, intense, collimated beam of $\nu_{e}$ or $\bar\nu_{e}$
through the beta decay of completely ionized radioactive ions. It
will be achieved by producing, collecting, and accelerating these
ions and then storing them in a  ring. This proposal is being
studied in depth and will take full advantage of the existing CERN
accelerator complex. The main future challenge lies in
building an intense proton driver and the hippodrome-shaped
decay ring which are essential for this programme.\\

For generating the $\nu$ and $\bar\nu$ beams, the commonly
examined ions are  $^{18}$Ne and $^{6}$He,
respectively~\cite{jacques}. The beams are almost free from all
types of {\em systematics}. The energy reach of these beta-beams
depend on the relativistic boost factor $\gamma$. Using the
existing CERN-SPS accelerator up to its maximum power, it will be
possible to achieve $\gamma\sim 250$~\cite{gamma_250}. A medium
$\gamma\sim 500$ beta-beam would require a refurbished SPS with
superconducting magnets or an acceleration technique utilizing the
LHC~\cite{burguet,gamma_250,LHC_upgrade}. A high $\gamma\sim 800$
could be achievable in the LHC era~\cite{gamma}.\\

The choice of ions for a beta-beam is predicated by the intended
physics. The low end-point energies (cf.  Table~1) of the $^{6}$He and
$^{18}$Ne ions restrict the energy reach of the beam; a threshold
energy of $3.5$ GeV is necessary to produce a $\tau$-lepton from an
incoming neutrino. Therefore, we choose to consider the recent
proposal~\cite{rubbia} of replacing the $^{6}$He and $^{18}$Ne ions by
$^{8}$Li and $^{8}$B ions respectively,
%%%%%%%%%%%%%%%%%%%%%%%%%%%%%%%%%%%%%%%%%%%%%%%%%%%%%%%%%%%%%%%%%%%%%%%%
\begin{table}\label{parameter}
\begin{center}
\begin{tabular}{|c|c|c|c|c|c|} \hline
   Ion & $t_{1/2}$ (s) & $E_0$ (MeV)& $f$& Decay fraction & Beam \\
\hline
  $^{18} _{10}$Ne &   1.67 & 3.41&820.4&92.1\%& $\nu_{e}$    \\
  $^6 _2$He   &   0.81 & 3.51&934.5&100\% &$\bar\nu_{e}$    \\
\hline
 $^{8} _5$B&0.77 & 13.92&600684.3&100\%&$\nu_{e}$    \\
 $^8 _3$Li&0.83 &12.96 &425355.2& 100\% & $\bar\nu_{e}$    \\
\hline
\end{tabular}
\caption{\em Half-life, end-point energy $E_0$, $f$-value
and decay fraction for various ions proposed for
beta-beams~\cite{beta}. In the presence of RPV couplings, $\beta$-decay of
these ions can give rise to neutrinos and anti-neutrinos of other
flavours with tiny branching ratios.}
\end{center}
\end{table}
%%%%%%%%%%%%%%%%%%%%%%%%%%%%%%%%%%%%%%%%%%%%%%%%%%%%%%%%%%%%%%%%%%%%%%%%
offering higher end-point energies. This option holds promise as it
has been shown~\cite{rubbia} that intense $^{8}$Li and $^{8}$B fluxes
can be obtained by the ionisation cooling technique. Higher $E_0$
implies that, for a given $\gamma$, the neutrino beams will be more
efficient in producing tau leptons. $^{8}$Li and $^{8}$B possess
similar half-life and $A/Z$ ratio as $^{6}$He and $^{18}$Ne,
respectively. So they share the same key characteristics for bunch
manipulation~\cite{0605101}. We pick only the neutrino beam for our
discussion.\\

The geometry of the beta-beam storage ring determines the neutrino
flux at a near-detector. For a low-$\gamma$ design, a 6880 m decay
ring with straight sections of length ($\equiv S$) $2500$ m
 each ($36\%$ useful length for ion decays) has been
proposed. In such a configuration, $N_0=1.1\times 10^{18}$ useful
decays (decays in one of the straight sections) per year can be
obtained with $^{18}$Ne ions~\cite{autin,terr}. We have used this
same luminosity for $^{8}$B and higher $\gamma$~\cite{donini}. To
settle these issues a dedicated study is on at CERN.

%%%%%%%%%%%%%%%%%%%%% FIGURE 1,  DETECTOR DESCRIPTION %%%%%%%%%%%%%%%%%%
\begin{figure}[thb]
\hskip 1.35cm
\psfig{figure=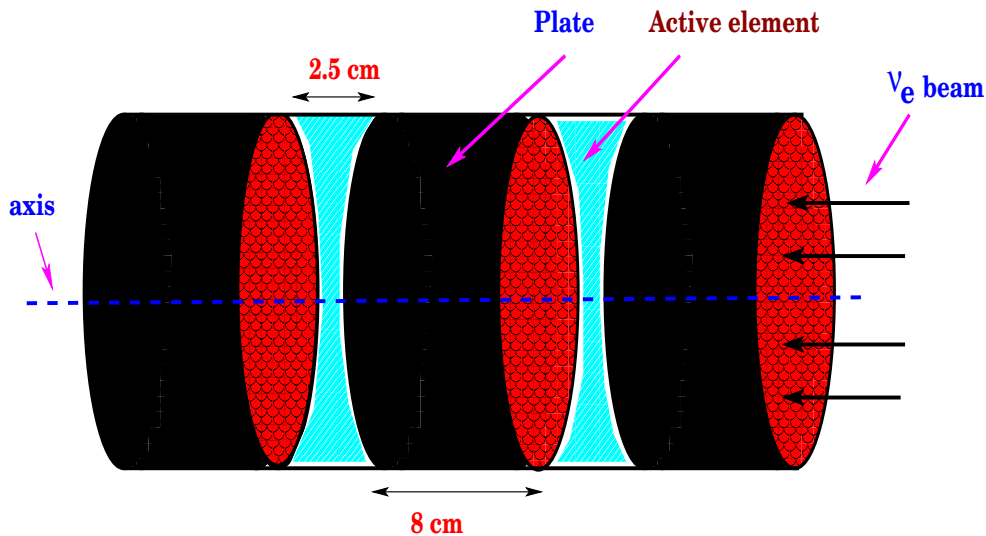,width=13.0cm,height=7.0cm,angle=0}
\caption{\sf \small A schematic diagram of the proposed
detector (a part only). The incoming $\nu_e$ beam may have a very small
contamination of neutrinos of other flavours in the presence of lepton
flavour violating interactions. }\label{fig1} \vskip 0.5cm
\end{figure}
%%%%%%%%%%%%%%%%%%%%%%%%%%%%%%%%%%%%%%%%%%%%%%%%%%%%%%%%%%%%%%%%%%%%
\subsection{Detector Simulation Study}

We consider a cylindrical 5 kT detector (as in Fig.~\ref{fig1})
aligned with one of the straight sections of the storage ring.  The
detector is made of iron slabs (thickness 8 cm) with interleaved
active detector elements (thickness 2.5 cm) such as resistive plate
chambers (RPCs). The readouts from these RPCs will be concentric
annular strips of small width with further segmentation to improve the
position resolution. In this proposal, iron is the main content of the
detector\footnote{Lead may be an interesting alternative material to
enhance the event rate.}. The thickness of the slabs ensures that
electrons do not propagate in the detector.  The signal muons are of
sufficient energy to give rise to long tracks. To eliminate possible
beam-induced backgrounds (see below) from pions produced in charged
and neutral current processes, typically 6 to 13 hits (depending on
the boost $\gamma$ ranging from 250 to 450) are required of a putative
muon track.\\

As noted earlier, the signature of new physics we consider is the
appearance of prompt tau leptons which decay into muons with a
branching fraction of $17.36\%$~\cite{hagiwara}. The tau
production threshold is around $3.5$ GeV. This is what necessitates
the higher boost $\gamma$. \\

Backgrounds, other than those of the beam-induced variety
discussed below, are controllable, as we now point out. A
beam-off run will help make a first estimate of these
backgrounds. Further,  an important aspect of the beta-beam
source is its capability of eliminating backgrounds through
timing information. The beam itself will consist of bunches of
typically 10ns size and the number of bunches will be chosen so
as to ensure that the ratio of the active- to the dead-time is
$\cal{O}$($10^{-3}$)~\cite{timing}.  Backgrounds from other
sources, namely, atmospheric neutrinos, spallation neutrons,
cosmic rays, etc. can thus be largely rejected from the
time-stamp of a recorded event.  Even further reductions of the
backgrounds of external origin can be envisioned through 
fiducial and directionality cuts.\\

Now let us turn our attention to the issue of beam-induced backgrounds
caused by neutral and charged current interactions of unoscillated
$\nu_e$. Electrons produced through weak interactions by the incoming
$\nu_{e}$ are quickly absorbed and do not leave any track. Formation
of prompt muons through $R$-parity violating supersymmetric
interactions is suppressed by strong bounds on the relevant couplings
arising from limits on $\mu-e$ transitions in
atoms~\cite{mu2e}. However, the beta-beam neutrinos can produce pions
along with other hadrons at the detector via charged current and
neutral current processes. They undergo strong interactions with the
detector material and are quickly absorbed before they can decay. But
as numerous pions are produced, it needs to be checked whether some of
them can fake the signal.\\

We have checked our na\"{i}ve expectations with a detector simulation
study using GEANT~\cite{geant3} aided by NUANCE~\cite{nuance}. We
observe that for neutrino-nucleon interactions at energies interesting
for our study, the produced lepton preferentially carries most of the
energy of the incident neutrino. Moreover, pions are usually produced
with multiplicity more than unity. Hence it is not unreasonable to
expect that the pions will be less energetic compared to the taus
produced via $\not{\! L}$ interactions and hence in detectors of this
genre, it is possible to distinguish hadronic showers from a muon
track.\\

However, we followed a conservative approach in pion background
estimation. Although pions do not leave behind a straight track
like a muon, we still count the number of hits as a measure of
the distance traversed by a pion. We impose a criterion of
minimum number of hits to identify a track to be a muon one. We
find from a simulation that, for $\gamma=250/350/450$, imposing a
cut of 6/10/13 hits will reduce the pion background at least to
the $10^{-3}$ level.\\

\begin{figure}[hbt]
\begin{center}
\hskip 0.4cm
\psfig{figure=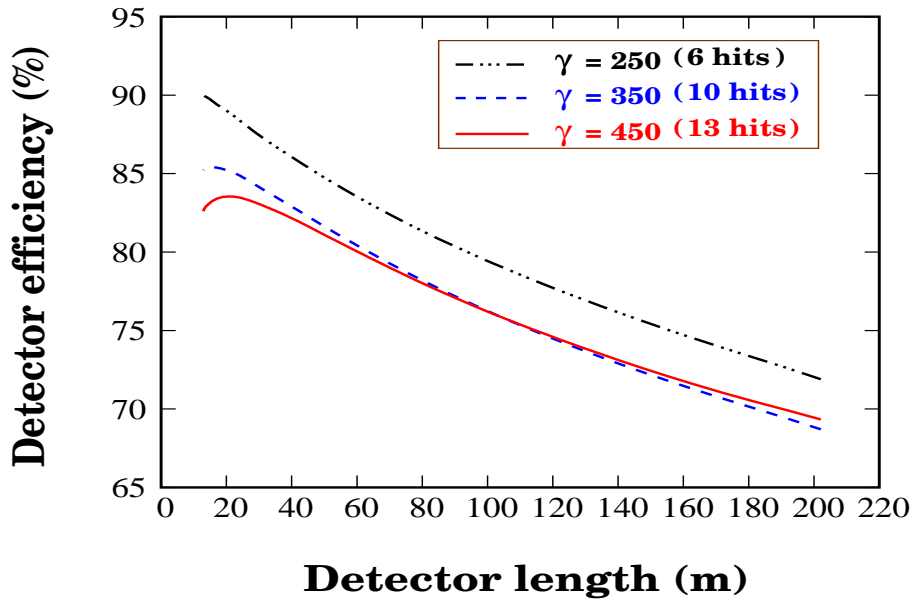,width=12.0cm,height=8.0cm,angle=0}
\caption{\sf \small Detector efficiency for
$\gamma=250,350,450$. The corresponding cuts on muon hits used are
6, 10 and 13 respectively.} \label{efficiency}
\end{center}
\end{figure}

The detector geometry plays a role in determining the signal
efficiency after imposition of these cuts. Since the muons produced
from boosted tau lepton decay carry transverse momentum, some of them
may exit the detector through its sides, failing to satisfy the cuts.
For a fixed detector mass (5 kT), a longer detector has a smaller
cross-sectional area, resulting in a drop in the detector efficiency
for the above reason.  As the detector length increases from 20 m to
200 m, with our set of cuts, the efficiency factor reduces from ~85\%
to 70\% approximately, showing little dependence on $\gamma$ (see
Fig.~\ref{efficiency}).

%%%%%%%%%%%%%%%%%%%% FIGURE Efficiency%%%%%%%%%%%%%%%%%%%%%%%%%%%%%%
%%%%%%%%%%%%%%%%%%%%%%%%%%%%%%%%%%%%%%%%%%%%%%%%%%%%%%%%%%%%%%%%%%%%%

\subsection{Neutrino fluxes}

Neglecting small Coulomb interactions, the lab frame neutrino
beta-beam flux (per unit solid angle per unit energy bin per unit
time per unit length of the straight section) emitted at an angle
$\theta$ with the beam axis is described by~\cite{agarwalla1}
\beq
 \phi(\en,\theta) = \frac{1}{4\pi}\frac{g}{m_e^5 \,f} \: \frac{1}{\gamma(1-\beta
 \cos\theta)}\: E_e^{\star}\: (\ens)^2 \: \sqrt{(E_e^{\star})^2-m_e^2}
\eeq
where $E_e^{\star}(\equiv E_0+m_e-\ens)$, and $\ens(\equiv \en \gamma
(1-\beta\cos\theta))$ are the rest frame energies of the emitted
electron and the neutrino\footnote{Quantities without the `$\ast$'
refer to the lab frame.}. $g \equiv N_0/S$ is the number of useful decays
per unit time per unit length of the straight section. $m_e$
represents the electron mass.
$f$ and $E_0$ refer to the decaying ion as listed in Table~1.\\

To calculate the resulting number of events at a cylindrical
near-detector of radius $R$ and length $D$ aligned with the beam
axis it is necessary to integrate over the length $S$ of the
straight section of the storage ring and the volume of the
detector. The event rate at a detector placed at a distance $L$
from the storage ring is given by~\cite{volpe}
\beq \frac{dN_{\not L}}{dt}
 =n \varepsilon \int_0^S \: dx \int_0^D \: d\ell
\int_0^{\theta'} d\theta \:{2\pi} \sin\theta
\int_{E_{\nu}^{min}}^{E_{\nu}^{\prime}} \: d\en \:\phi(\en,\theta)\:
\sigma(\en), \eeq
where
\begin{equation}
\tan\theta^{'}(x,\ell)=\frac{R}{L+x+\ell}\,~~{\rm
and}~~\en^{\prime}=\frac{E_0}{\gamma(1-\beta\cos\theta)}.
\end{equation}
Here $n$ represents the number of target nucleons per unit
detector volume, $\varepsilon$ is the detector efficiency as
presented in Fig.~\ref{efficiency}, $\en^{min}$ denotes the tau
production threshold, and $\sigma(\en)$ stands for the
neutrino-nucleon cross section.  Note that the source of
$L$-violation may lie either in $\phi(\en,\theta)$ (in case of RPV
$\beta$-decay) or in
$\sigma(\en)$ (in case of RPV tau production).\\

To help subsequent discussion, following ~\cite{volpe}, we
rewrite the above formula isolating the geometry integrated total
flux $\Phi(\en;S,D,R,L)$ (per unit time per unit energy bin)
falling on the detector and emitted from the whole length of the
straight section as follows: %
\beq \frac{dN_{\not L}}{dt} =n \varepsilon
\int_{E_{\nu}^{min}}^{E_{\nu}^{max}} \:\: d\en \Phi(\en;S,D,R,L)\:
\sigma(\en), \eeq
where
\beq \Phi(\en;S,D,R,L)=\int_0^S \: dx \int_0^D \: d\ell
\int_0^{\theta'} d\theta \:{2\pi} \sin\theta \:
\phi(\en,\theta)\label{PHIeqn}\eeq
and
\beq \en^{max}=\frac{E_0}{\gamma(1-\beta)}. \eeq
The beta-beam also involves a few small uncertainties which we neglect
in our analysis. However for completeness, we list them here:
\begin{itemize}
\item There exist different excited states of the daughter nuclei
of the decaying ion, which additionally lead to small
contributions to the spectra with different endpoint energies.

\item The ion beam has a finite transverse size. However, as this
size varies~\cite{adrian} between only $3.0$ cm to $5.1$ cm, with
an average of $4$ cm ($3\sigma$), in both transverse directions
inside the ring, the variation in flux at the detector due to this
is negligible.

\item The decaying ions may have small transverse momentum due to
thermal fluctuations ($k_{B}T \sim 2.6\times 10^{-3}$ eV), but this
can be safely ignored in comparison with the end-point energy of the beta
decay.

\end{itemize}

\section{$L$ violating processes}\label{sec3}

Lepton number violation arises naturally as one supersymmetrises
the standard model. In the minimal supersymmetric standard model,
lepton number and baryon number ($B$) conservation is ensured by
invoking `$R$-parity'. It is a discrete $Z_2$ symmetry under
which the SM particles are even and their superpartners are odd.
The imposition of such a symmetry, while it serves a purpose, is
rather {\em ad hoc}. In general, from the na\"{i}ve
theoretical point of view it is expected that $L$ and $B$
conservation does not hold in supersymmetric theories. However,
as this leads to a very fast proton decay, we follow a common
practice and assume that $B$ is conserved. This can be ensured by
replacing the $Z_2$ symmetry of $R$-parity by a $Z_3$ symmetry,
the so-called `baryon triality'~\cite{ross}. In such a scenario,
in addition to the usual Yukawa interactions, the superpotential
contains renormalizable $L$-violating trilinear $\lambda$- and
$\lambda'$-type couplings and bilinear $\mu_i$ couplings:
\begin{equation}
W_{\not L} = \frac{1}{2}\lambda_{ijk} {L}_i {L}_j {E}_k^c +
\lambda_{ijk}' {L}_i {Q}_j {D}_k^c + \mu_i {L}_i {H}_u,
\label{super}
\end{equation}
where $i,j,k = 1, 2, 3$ are generation indices.
Here ${L}_i$ and ${Q}_i$ are $SU(2)$-doublet lepton and quark
superfields respectively; ${E}_i$, ${D}_i$ denote the
right-handed $SU(2)$-singlet charged lepton and down-type quark
superfields respectively; ${H}_u$ is the Higgs superfield which
gives masses to up-type quarks. $\lambda_{ijk}$ is antisymmetric
under the interchange of the first two generation indices.  The
bilinear couplings, $\mu_i$, are severely constrained by the
small neutrino masses. So we will discuss the phenomenology of
$\lambda$ and $\lambda'$ type couplings only. Then, the above
superpotential leads to the following Lagrangian:
\begin{eqnarray}
{\cal {L_{\not L}}} &=&  \lambda'_{ijk} ~\big[ ~\tilde d^j _L
\,\bar d ^k _R \nu^i _L
  + (\tilde d ^k_R)^\ast ( \bar \nu ^i_L)^c d^j _L +
   \tilde \nu ^i _L \bar d^k _R d ^j _L  \nonumber \\
& & ~~~~~ -\tilde e^i _L \bar d ^k _R u^j _L - \tilde u^j _L
\,\bar d ^k _R e^i _L
-(\tilde d^k _R)^\ast (\bar e ^i _L)^c u^j _L \big] \nonumber \\
&& +\frac{1}{2}\lambda_{ijk} ~\big[ ~\tilde e^j _L \,\bar e^k_R
\nu^i_L
  + (\tilde e ^k_R)^\ast (\bar \nu ^i _L)^c e^j _L +
   \tilde \nu ^i _L \bar e^k _R e ^j _L  - (i \leftrightarrow j)
\big] + h.c. \label{lag_2}
\end{eqnarray}
The above interaction terms violate lepton number, $L$, as well as
lepton flavour number. Suitable combinations of two such terms can
lead to processes which are lepton flavour violating but
$L$-conserving. The study of such non-standard interactions at a
neutrino factory has been undertaken in~\cite{datta,huber}.
Influence of these interactions in the context of long baseline
beta-beam experiments was studied in~\cite{agarwalla2}.  Here we
examine the physics potential of beta-beams to explore such
interactions in a near-detector scenario. To impose conservative
upper bounds, we work in a minimal RPV framework where only a pair
of such couplings are assumed to be non-zero at a time.\\

For a near-detector, RPV can come into effect in
two ways as described in the following subsections.

\subsection{RPV and $\nu_\tau$ production in  $\beta$-decay}

\begin{figure}[tbh]
\begin{center}
\epsfig{file=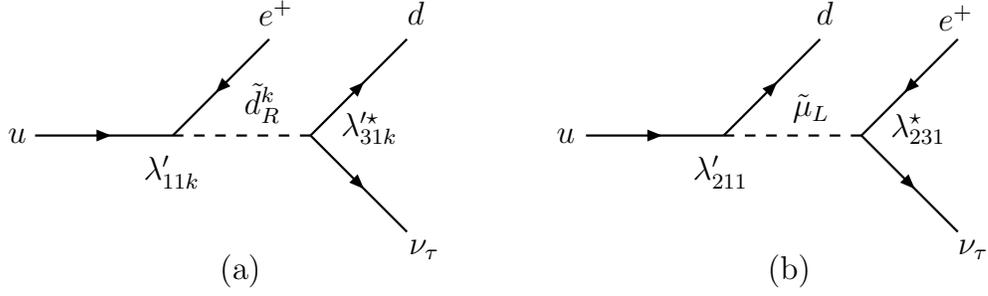,width=13cm}
%%%%%%%%%%%%%%%%%%%%%%%%%%%%%%%%%%%%%
\vskip0.8cm \caption{ {\sf \small Feynman diagrams for RPV driven
$\beta$-decay through (a) $\lambda'\lambda'$ and (b) $\lambda\lambda'$
type trilinear product couplings. Substantial event rates are obtained
in (a) when $k=2,3$.} } \label{Feyn} \vskip0.5cm
\end{center}
\end{figure}

RPV interactions can drive beta decay producing $\nu_\tau$ instead of
$\nu_e$ (see Fig.~\ref{Feyn}(a)). $\nu_\tau$ so produced 
give rise to $\tau$ leptons in the detector which
may decay in the leptonic channel producing muons.\\

%%%%%%%%%%%%%%%%%%%%%%%%%  FIGURE 3 %%%%%%%%%%%%%%%%%%%%%%%%%%%%%%%%%%%%
%%%%%%%%%%%%%%%%%%%%%%%%%%%%%%%%%%%%%%%%%%%%%%%%%%%%%%%%%%%%%%%%%%%%%%%%
Simultaneous presence of ${\lambda}'_{31k}$ and ${\lambda}'_{11k}$
couplings can be responsible for producing a $\nu_\tau$ in 
$\beta$-decay. Of these, $\lambda'_{111}$ is tightly constrained from
neutrinoless double beta decay~\cite{double_beta}. But the upper bound
on the combination $|{\lambda'^{\star}_{31k}
\lambda'_{11k}}|, k=2,3$ is rather relaxed; a limit of
$2.4\times 10^{-3} ({\tilde m}/{100~{\rm GeV}})^2$, $\tilde m$ being a
common sfermion mass, follows from $\tau^{-} \rightarrow e^{-}
\rho^{0}$~\cite{mu2e,hagiwara}.  $\tilde m$ denotes a common sfermion
mass. The corresponding decay amplitude can be written as,
%%%%%%%%%%%%%%%%%%%%%%%%%%%%%%%%%%%%%%%%%%%%%%%%%%%%%%%%
\begin{eqnarray}
{M}_{\not L}(u \longrightarrow d e^+ \nu_\tau) &=&
\frac{\lambda'^{\star}_{31k}\lambda'_{11k}}
{2 (\hat s - {\tilde m}^2)}\; \big[\bar u_{\nu_\tau} \gamma_{\mu} P_L
u_{e}\big]\, \big[\bar u_{d} \gamma^{\mu} P_L u_u \big].
\label{process01}
\end{eqnarray}
\noindent
%%%%%%%%%%%%%%%%%%%%%%%%%%%%%%%%%%%%%%%%%%%%%%%%%%%%%%%%
Alternatively, $\nu_\tau$ can be produced in $\beta$-decay if another
combination of RPV couplings ${\lambda^\star_{i31}\lambda'_{i11}}$
($i$ = 1,2) is non-zero (see Fig.~\ref{Feyn}(b)).  As mentioned
earlier, $\lp_{111}$ is severely constrained. The combination
$|{\lambda^\star_{231}
\lambda'_{211}}|$ is bounded from above  by $1.6\times 10^{-3}
({\tilde m}/{100~{\rm GeV}})^2$ arising from the decay channel
$\tau^{-} \rightarrow e^{-} \eta^{0}$~\cite{mu2e,hagiwara}, which
is not too small to produce an observable effect. The
corresponding decay amplitude is given by, 
%%%%%%%%%%%%%%%%%%%%%%%%%%%%%%%%%%%%%%%%%%%%%%%%%%%%%%%%
\begin{eqnarray}
{M}_{\not L}(u \longrightarrow d e^+ \nu_\tau) &=&
\frac{\lambda^\star_{231}\lambda'_{211}}
{(\hat t - {\tilde m}^2)}\;
 \big[\bar
u_{\nu_\tau} P_R \,u_{e}\big]\, \big[\bar u_{d} P_L u_u \big].
\label{process02}
\end{eqnarray}
\noindent
%%%%%%%%%%%%%%%%%%%%%%%%%%%%%%%%%%%%%%%%%%%%%%%%%%%%%%%%

\subsection{RPV in tau production from $\nu_e$}

$\nu_e$ produced through ordinary $\beta$-decay driven by weak
interactions can undergo RPV interactions with the detector producing
$\tau$ which subsequently decay into muons. \\

%%%%%%%%%%%%%%%%%%%%%%%%%  FIGURE 4 %%%%%%%%%%%%%%%%%%%%%%%%%%%%%%%%%%%%
\begin{figure}[hbt]
%%%%%%%%%%%%%%%%%%%%%%%%%%%%%
\begin{center}
\epsfig{file=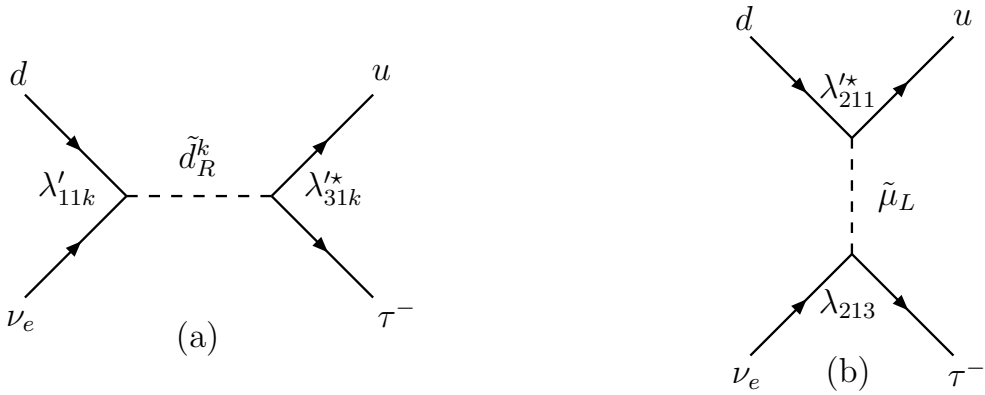,width=13cm}
%%%%%%%%%%%%%%%%%%%%%%%%%%%%%%%%%%%%%
\vskip0.8cm \caption{ {\sf \small Feynman diagrams for tau
production from an incoming $\nu_{e}$ beta-beam through (a)
$\lambda'\lambda'$ and (b) $\lambda\lambda'$ type trilinear
product couplings. Substantial event rates are obtained in (a)
when $k=2,3$.} } \label{Feyn2} \vskip0.5cm
\end{center}
\end{figure}
%%%%%%%%%%%%%%%%%%%%%%%%%%%%%%%%%%%%%%%%%%%%%%%%%%%%%%%%%%%%%%%%%%%%%%%%
Simultaneous presence of ${\lambda}'_{31k}$ and ${\lambda}'_{11k}$
couplings can give rise to $\tau^{-}$ in the final state from an
incoming $\nu_{e}$ of the beta-beam (see Fig.~\ref{Feyn2}(a)).  The
amplitude for the corresponding $s$-channel diagram can be written,
after a Fierz transformation, as
%%%%%%%%%%%%%%%%%%%%%%%%%%%%%%%%%%%%%%%%%%%%%%%%%%%%%%%%
\begin{eqnarray}
{M}_{\not L}(\nu_e \;d \longrightarrow \tau^- \; u) &=&
\frac{\lambda'^{\star}_{31k}\lambda'_{11k}}
{2 (\hat s - {\tilde m}^2)}\; \big[\bar u_{\tau} \gamma_{\mu} P_L
u_{\nu_e}\big]\, \big[\bar u_{u} \gamma^{\mu} P_L u_d \big].
\label{process1}
\end{eqnarray}
\noindent
%%%%%%%%%%%%%%%%%%%%%%%%%%%%%%%%%%%%%%%%%%%%%%%%%%%%%%%%

An alternative channel of tau production from an incoming $\nu_e$ beam
exists (see Fig.~\ref{Feyn2}(b)) if a particular combination of the
$\lambda$ and $\lambda'$ couplings
${\lambda_{i13}\lambda'^\star_{i11}}$ ($i$ = 2,3) is non-zero.  Here
again, $\lambda_{313}$ is severely constrained from neutrinoless
double beta decay experiments~\cite{Bhattacharyya}. An upper bound of
$1.6\times 10^{-3} ({\tilde m}/{100~{\rm GeV}})^2$ applies to the
combination $|{\lambda_{213} \lambda'^\star_{211}}|$, from the decay
channel $\tau^{-} \rightarrow e^{-}
\eta^{0}$~\cite{mu2e,hagiwara}. The amplitude for this $t$-channel
process is
%%%%%%%%%%%%%%%%%%%%%%%%%%%%%%%%%%%%%%%%%%%%%%%%%%%%%%%%%%%%%%%
\begin{eqnarray}
{M}_{\not L}(\nu_e \;d \longrightarrow \tau^- \; u) &=&
\frac{\lambda_{213}\lambda'^\star_{211}} {(\hat t - {\tilde
m}^2)}\;
 \big[\bar
u_{\tau} P_L u_{\nu_e}\big]\, \big[\bar u_{u} P_R u_d \big].
\label{process2}
\end{eqnarray}
\noindent
%%%%%%%%%%%%%Feynman diagrams%%%%%%%%%%%%%%%%%%%%%%%%%%%%%%%%%%%%%%%%%%%
In what follows, we categorise the above two kinds of diagrams (a) and
(b) in Figs. 2 and 3 as $\lambda'\lambda'$ and $\lambda \lambda'$
processes, respectively.\\

Note that, if $|{\lambda'^{\star}_{31k} \lambda'_{11k}}|, k=2,3$ is
non-zero, tau leptons can be produced at the detector either due to
RPV interactions affecting $\beta$-decay or due to RPV interactions of
a $\nu_e$ with the detector material. These two equal contributions
add in the total rate of tau production.\\

However, for the $\lambda\lp$ process, we see that the RPV
combinations $|{\lambda^\star_{231} \lambda'_{211}}|$ (which drive the
RPV beta decay) and $|{\lambda_{213}
\lambda'^\star_{211}}|$ (which is responsible for producing a tau
from an incoming $\nu_e$ in the detector) are different. As we are
following the strategy of taking only two RPV couplings non-zero at a
time, these contributions, which are of the same magnitude, cannot be
present at the same time. \\

In passing, a few comments are in order:
\begin{itemize}
\item In both diagrams, the incoming $\nu_e$
can interact with a $\bar u$ quark from the sea to
produce a tau. Due to the smallness of the corresponding parton
distribution function, this contribution is suppressed
but we do include it in the numerical evaluations.

\item Here we should mention that the flavour changing neutral
current process, $K^{+} \rightarrow \pi^{+} \nu \bar\nu$
\cite{kpipi} puts stringent bounds on all the $\lambda'$
couplings. However, these are basis dependent and hence can be
evaded.

\item As already noted, the non-observation of the process $\mu
\rightarrow e (Ti)$ severely restricts~\cite{latest_bounds,mu2e}
the possibility of emitting a $\nu_\mu$ in $\beta$-decay and
direct production of muons from an incoming
$\nu_e$ beam.

\item Since the beta-beam energy is $\sim$ a few GeV, the expected
event rate will be essentially independent of the sfermion mass as
the bounds on $\lambda, \lambda^\prime$ scale with $(\tilde
m/100~{\rm GeV})^2$.

\end{itemize}

At this energy range it is important to consider contributions from
deep-inelastic, quasi-elastic, and single-pion production channels. To
estimate the RPV deep-inelastic scattering cross section, we have used
CTEQ4LQ parton distributions~\cite{parton}.  RPV quasi-elastic
scattering and single-pion production cross sections have been
evaluated from the corresponding SM cross sections\footnote{These
cross sections include all nuclear effects for an iron
target.}~\cite{paschos} by a rescaling of the couplings. We have
noticed that, as eq.~\ref{process2} is not reducible to an SM-like
$(V-A) \otimes (V-A)$ Lorentz structure, in calculating DIS cross
section a factor $\sim 1/3$ appears from polar integration compared to
that for eq.~\ref{process1}. For the $\lambda\lambda'$ process we have
adopted the same suppression factor for the SM quasi-elastic and
single-pion production cross sections as well.  Conservatively,
we assume that a similar suppression also applies to the case of
RPV $\beta$-decay. It bears stressing that the effect of the
tau mass is felt on the neutrino-nucleus cross section throughout the
energy range beyond the $\tau$-threshold and this is included in the
analysis.

\section{Results}\label{sec4}

A near-detector setup is qualitatively different from a far-detector
as in the former case the storage ring and the detector really `sees'
each other and relative geometric considerations are of much
relevance. The observed number of events in a given period of time
depends on the choice of the radioactive ion, the boost factor
$\gamma$ and the details of the setup (which include storage ring
parameters, detector configuration and the short base-length between
them). As alluded to earlier, the maximum $\gamma$ available is
limited by the storage ring configuration. With a view to optimising
the setup, we summarise the essential inputs as follows:
\begin{itemize}
\item Storage ring parameters: Total length 6880 m, length of a
straight section, $S=2500$ m, number of beta decays in the
straight section, $N_0=1.1\times 10^{18}$ per year.

\item Detector configuration:  The detector material
is\footnote{Brief comments are made about a water \u{C}erenkov
detector in sec. IV.2.} iron ($\rho=7.87$ gm/cc). We consider a
detector of mass 5 kT. For a given material, this fixes
the length of the detector as the radius is changed. It varies from
202.13 m to 12.63 m as the radius ranges over 1m to 4m.

\item Base-length: Results are presented for three representative
values of the distance of the detector from the storage ring,
$L=200$ m, 500 m, 1 km.

\item Boost factor $\gamma$: The tau production threshold ($3.5$
GeV) calls for a high $\gamma$. We consider $\gamma=250, 350, 450$
for $^8$B and as large as $800$ for $^{18}$Ne.

\end{itemize}

\begin{figure}[hbt]
 \begin{center}
 %\vskip -1.15cm
\hskip 0.4cm
\psfig{figure=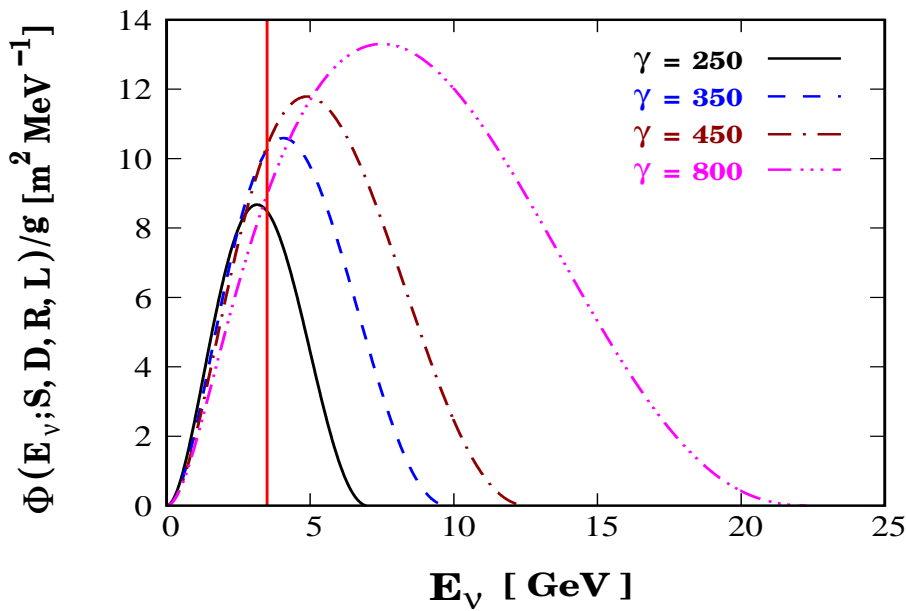,width=12.0cm,height=8.0cm,angle=0}
%\vskip -0.5cm
\caption{\sf \small Geometry integrated flux
$\Phi(\en;S,D,R,L)/g$  taking $^8$B as the decaying ion is
plotted against neutrino energy $\en$ for different $\gamma$ for
$S=2500$ m, $D=202.13$ m, $R=1$ m, and $L=200$ m. The vertical
line at 3.5 GeV indicates the tau production threshold energy.}
\label{capphi}
\end{center}
\end{figure}

The high collimation achievable in the beta-beams encourages the
choice of a detector of cylindrical shape coaxial with the storage
ring straight section. As  $\gamma$ increases, the RPV event rates
increase for the following reasons:
\begin{enumerate}
\item
 an increasingly larger part of the beam falls onto the detector,
\item
 more neutrinos have enough energy to produce a tau lepton,
\item
 with the more energetic neutrinos the cross section is
 larger.
\end{enumerate}
The
first two effects are demonstrated by Fig.~\ref{capphi}.%
%%%%%%%%%%%%%%%%%%%% FIGURE Phi  %%%%%%%%%%%%%%%%%%%%%%%%%%%%%%
%%%%%%%%%%%%%%%%%%%%%%%%%%%%%%%%%%%%%%%%%%%%%%%%%%%%%%%%%%%%%%%%%%%%%
The geometry integrated flux, $\Phi$, as defined in eq.~\ref{PHIeqn},
represents the beta-beam neutrino flux spectrum falling onto the
detector per unit time. It is seen that as $\gamma$ increases, the
total area under the curve also increases, illustrating the first
effect. The area under the curves on the right side of the vertical
line (the threshold) also increases with $\gamma$, in conformity with
the second expectation. For a high $\gamma$ the beam should
saturate. However, with the $\gamma$ used in Fig.~\ref{capphi} this is
not evident due to the enormous length of the straight section of the
storage ring: To collimate the flux emanating from the rear part of
the ring a very high $\gamma$ will be needed.\\

As geometry plays a crucial role in optimising the near-detector
setup, we study the detector length dependence of the expected number
of RPV events for different base-lengths and different $\gamma$. We
consider the contribution coming from the two options -- the
$\lambda'\lambda'$ and $\lambda\lambda'$ processes -- in different
panels for every figure, assuming the RPV coupling constants saturate
present experimental upper limits.\\

\subsection{Choice of ion source and detector}

The choice of $^8$B as the ion source provides the most attractive
option due to its high end-point energy. Iron calorimeters are
preferred for the smaller size and significant background removal.\\

To get a glimpse of the number of events one might expect in such
a setup, let us present the following estimate. A 5 kT Fe detector
of radius 1 m (length 202.13 m) placed at a distance 200 m from
the decay ring can give rise to 92 (24) muon events via the
$\lambda'\lambda'$ ($\lambda\lambda'$) process in 5 years
for\footnote{The corresponding
numbers for $\gamma = 350$ are 421 (103).} $\gamma=250$.\\

%%%%%%%%%%%%%%%%%%%%% FIGURE 6 (GAMMA VARIATION) %%%%%%%%%%%%%%%%%%%%%%%
\begin{figure}[hbt]
\vskip -0.15cm
\psfig{figure=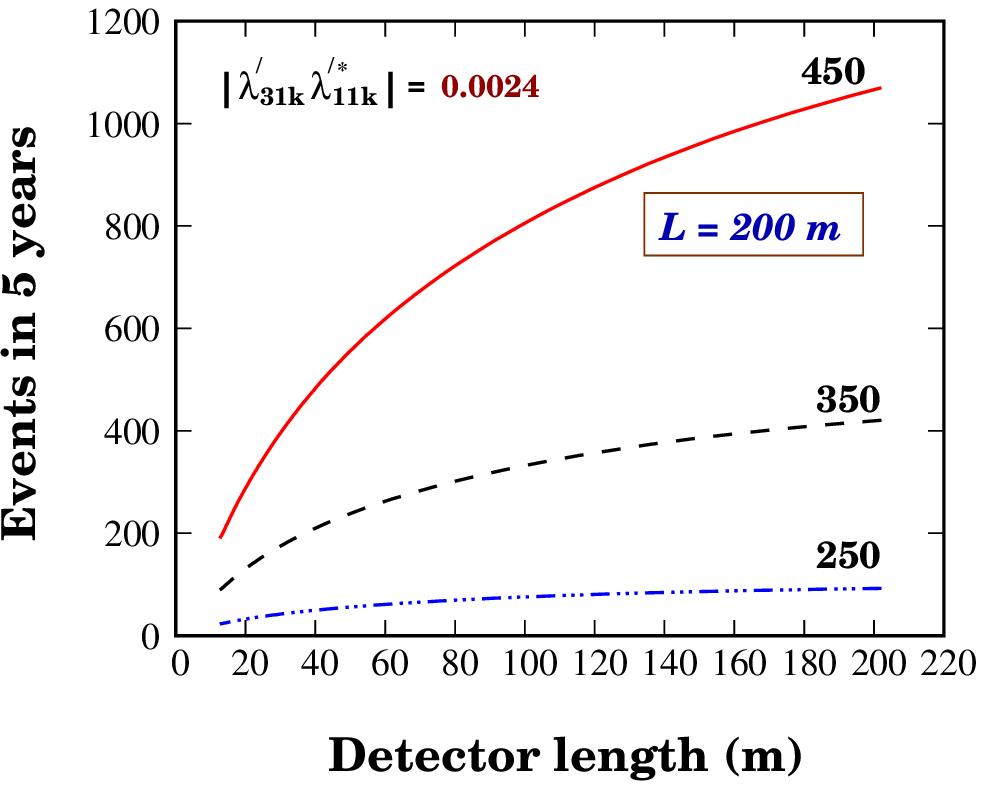,width=8.0cm,height=7.0cm,angle=0}
\caption{\sf \small Expected number of RPV muon events in five
years for a 5 kT iron detector vs. the detector length for
$\gamma$ = 250, 350, and 450 for $^8$B beta-beam flux. The
left~(right) panel is for the $\lp\lp$~($\lambda\lp$) driven
process. $k=2,3$. } \label{figgam} \vskip -8.9cm \hskip 8.3cm
\psfig{figure=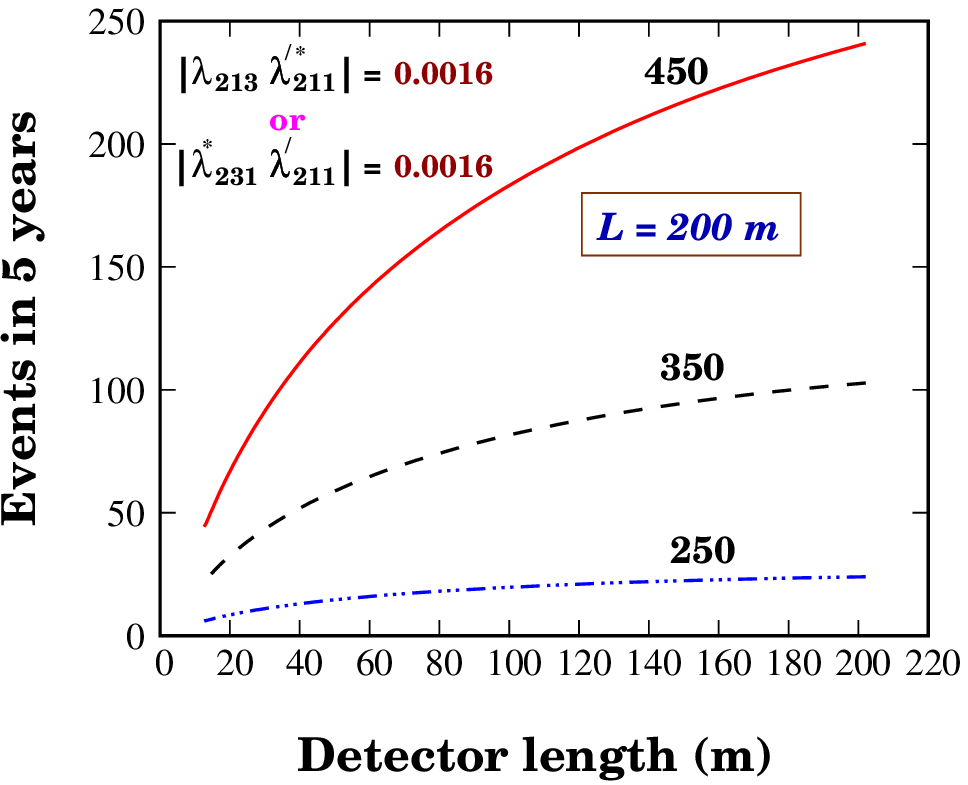,width=8.0cm,height=7.0cm,angle=0}
\vskip 1.5cm
\end{figure}
%%%%%%%%%%%%%%%%%%%%%%%%%%%%%%%%%%%%%%%%%%%%%%%%%%%%%%%%%%%%%%%%%%%%%
In Fig.~\ref{figgam} we exhibit the $\gamma$ dependence of the
expected number of muon events over a five-year period for a fixed
base-length of 200 m.  Collimation plays a role as is demonstrated
by the increase in the number of events for higher $\gamma$. As
expected, a long detector serves better as it provides more
opportunity for a neutrino interaction to occur. However, this
increase with the length is not linear; a part of the beam is lost
due to the concomitant decrease in the radius (to keep the total
mass fixed at 5 kT). In addition, with the increase in detector
length as the detector efficiency decreases, the increase
in the rates is also somewhat restricted. \\

\begin{figure}[hbt]
\hskip -0.15cm
\psfig{figure=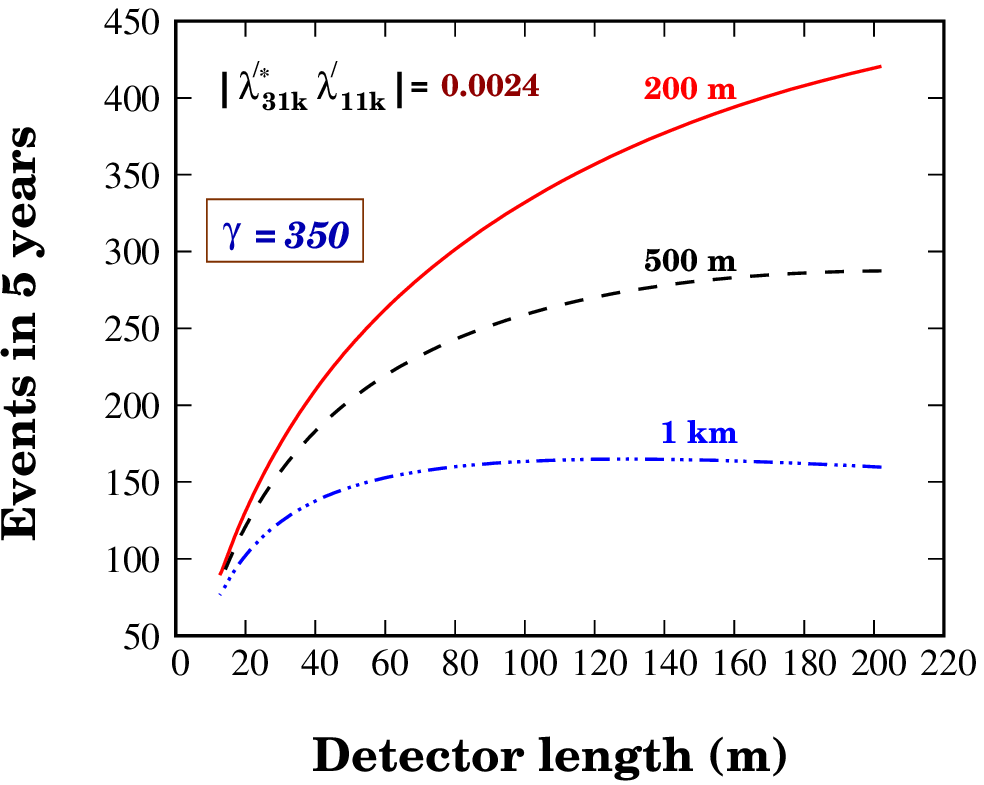,width=8.0cm,height=7.0cm,angle=0}
\caption{\sf \small Muon signal event rate in 5 years as a
function of the detector (Fe) length for three different choices
of base-length have been shown for $^8$B beta-beam flux. The
left~(right) panel corresponds to the $\lp\lp$~($\lambda\lp$)
driven process. $k=2,3$.}\label{figL} \vskip -8.9cm \hskip 8.3cm
\psfig{figure=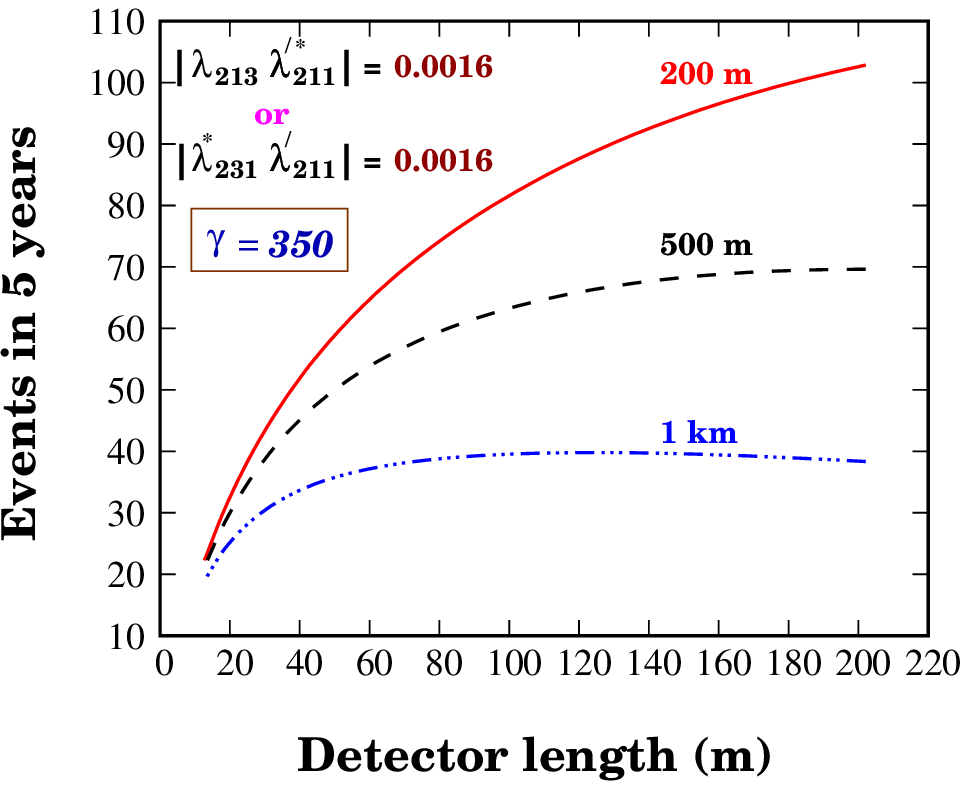,width=8.0cm,height=7.0cm,angle=0}
\vskip 2.0cm
\end{figure}

It is also of interest to study the base-length dependence of the
number of events. The beam spreads with an increase in the
base-length, reducing the effective flux hitting the detector.  This
causes a fall in the number of events (other parameters remaining the
same) as shown in Fig.~\ref{figL}. It is interesting to note that the
increase in the number of events with increase in the length of the
detector gets severely diluted at larger base-lengths. \\

%%%%%%%%%%%%%%%%%%%%% FIGURE 7  (BASELINE VARIATION) %%%%%%%%%%%%%%%%%%%
%%%%%%%%%%%%%%%%%%%%%%%%%%%%%%%%%%%%%%%%%%%%%%%%%%%%%%%%%%%%%%%%%%%%%

\begin{figure}[bht]
\hskip -0.15cm
\psfig{figure=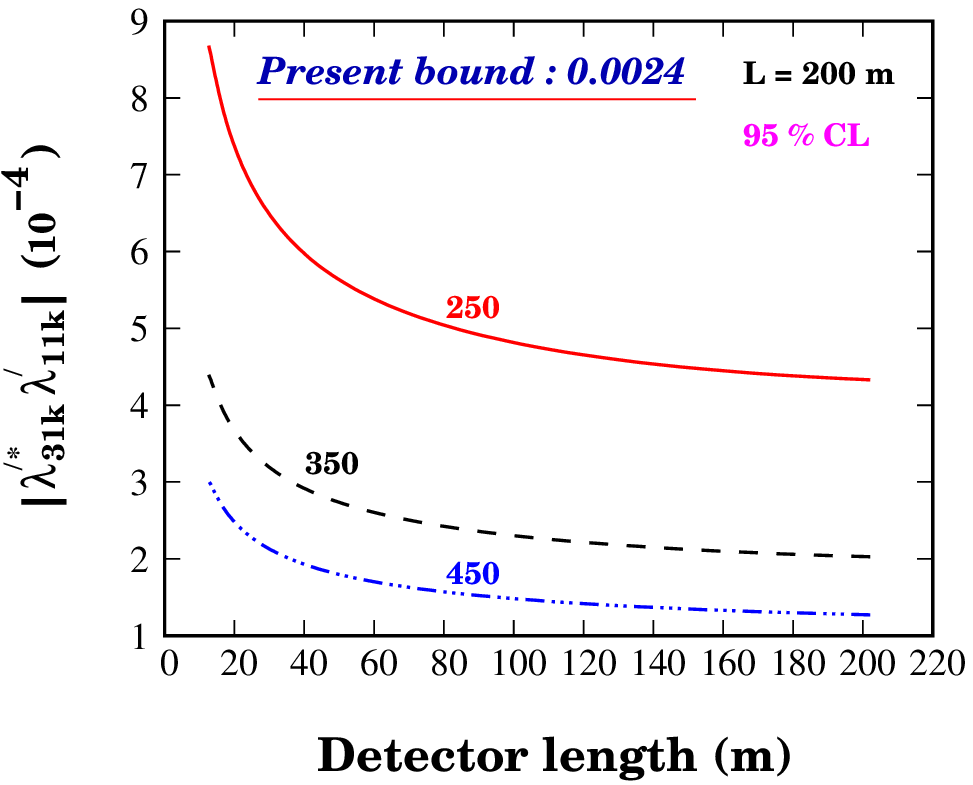,width=8.0cm,height=7.0cm,angle=0}
\caption{\sf \small Bounds on
$|\lambda'^{\star}_{31k}\lambda'_{11k}|,$ $k=2,3$
($|\lambda^\star_{231}\lambda'_{211}|$ or
$|\lambda_{213}\lambda'^{\star}_{211}|$) versus detector size at
$95\%$ CL for zero observed events is depicted in left (right) panel
for $\gamma=250,350,450$.  The bounds scale as $(\tilde m/100~{\rm
GeV})^2$. The results are for a five-year run for a 5 kT Fe detector
placed at a distance of 200 m from the front end of the storage ring
for $^8$B beta-beam flux.}\label{figbound} \vskip -9.33cm \hskip 8.3cm
\psfig{figure=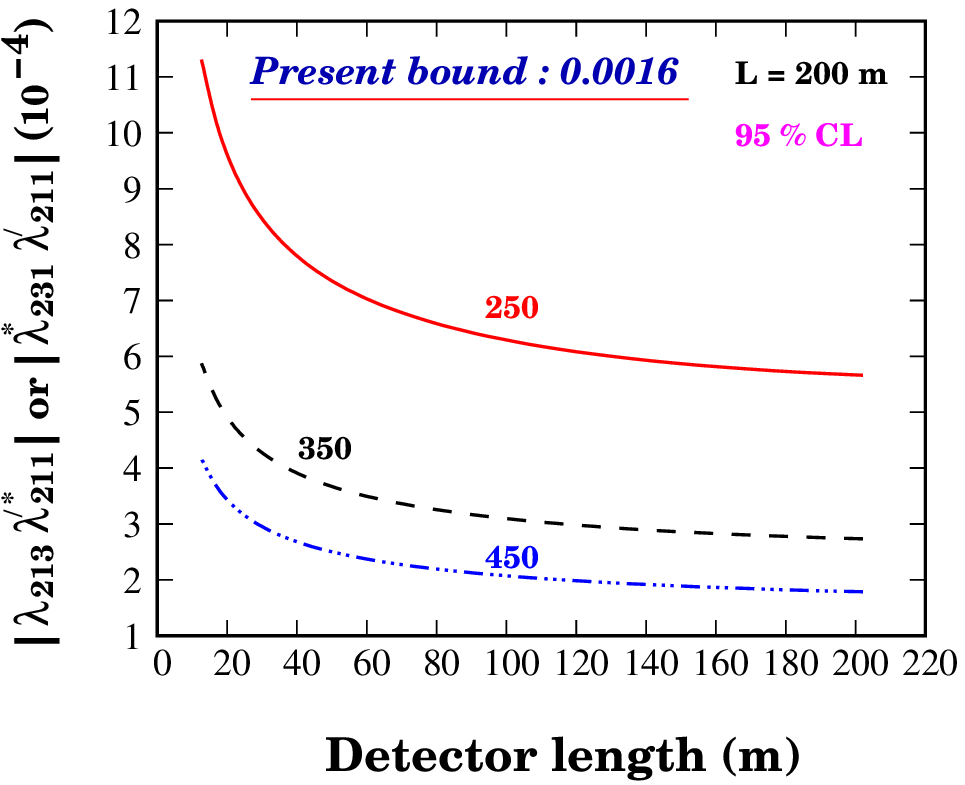,width=8.0cm,height=7.0cm,angle=0}
\vskip 2.0cm
\end{figure}

While presenting the expected number of events we assumed the RPV
couplings saturate the present experimental upper bounds.  In case
less or even no events are seen, the existing limits on the
combinations $|\lambda'^{\star}_{31k}\lambda'_{11k}|, k=2,3$,
$|\lambda^{\star}_{231}\lambda'_{211}|$ and
$|\lambda_{213}\lambda'^{\star}_{211}|$ will be improved. Choosing the
minimum number of non-zero RPV couplings, one can put conservative
upper bounds. In Fig.~\ref{figbound} we show the bounds -- the region
above the curves are disallowed -- achievable in the case of
`no-show'\footnote{At 95\% CL this corresponds to not more than 3
events.}. It is seen that to put stringent bounds it is necessary to
go for a higher $\gamma$ and a longer detector.

%%%%%%%%%%%%%  FIGURE 8  BOUND ON PROCESS 1 AND PROCESS 2 %%%%%%%%%%%%%%
%%%%%%%%%%%%%%%%%%%%%%%%%%%%%%%%%%%%%%%%%%%%%%%%%%%%%%%%%%%%%%%%%%%%%

\subsection{Alternative setups}

\begin{figure}[hbt]
\hskip -0.15cm
\psfig{figure=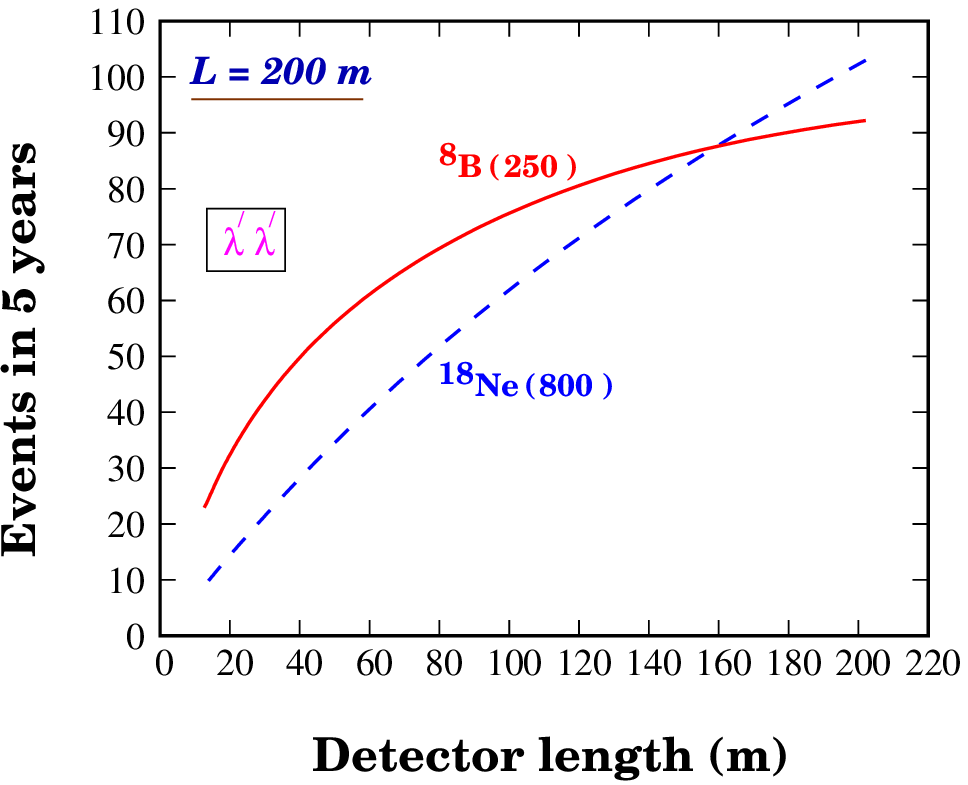,width=8.0cm,height=7.0cm,angle=0}
\caption{\sf \small Comparison of the muon signal event rates as a
function of the detector length for a 5 kT iron calorimeter
placed at a distance of 200 m from the storage ring
for $\gamma=800$~$(250)$ with $^{18}$Ne ($^8$B).
The left and right panels correspond to $\lp\lp$
and $\lambda\lp$ driven processes, respectively.}\label{figNe}
\vskip -8.9cm \hskip 8.3cm
\psfig{figure=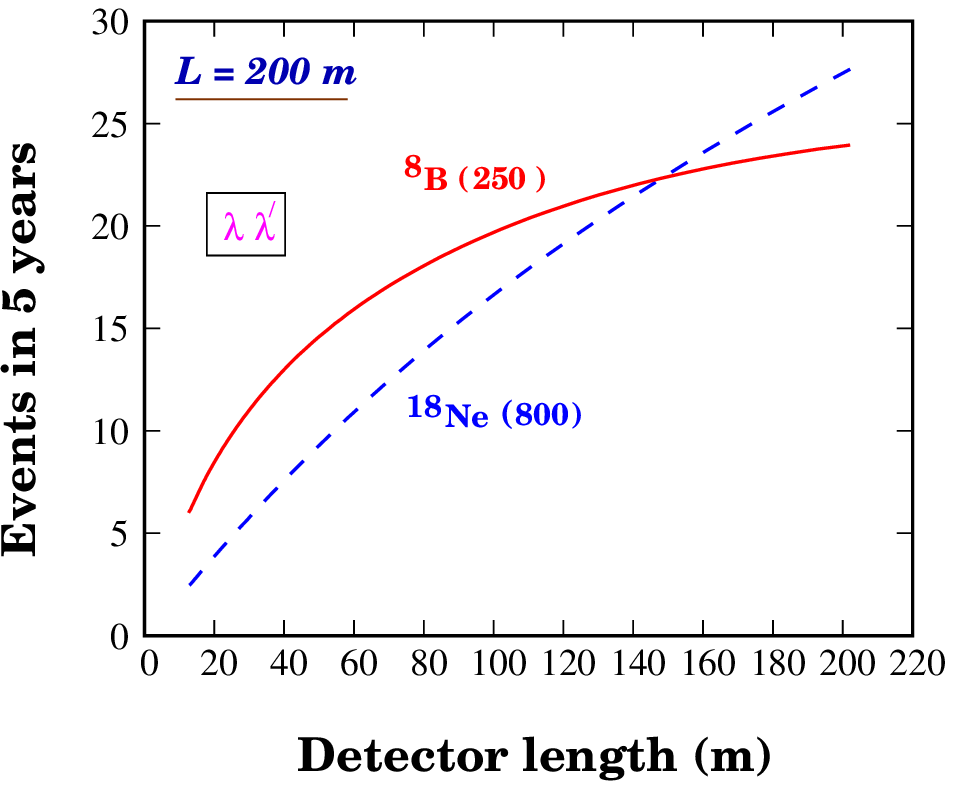,width=8.0cm,height=7.0cm,angle=0}
\vskip 2.0cm
\end{figure}

Although so far we have presented results with $^8$B as the
beta-beam source, $^{18}$Ne is the most discussed decaying ion in
the literature. As mentioned earlier, due to the smaller
end-point energy of $^{18}$Ne, a high $\gamma$ is required to
cross the $\tau$ threshold.  Fig.~\ref{figNe} depicts the
variation in the expected event rate with detector length for
$^{18}$Ne with $\gamma=800$ using a 5 kT iron calorimeter. We see
that due to high $\gamma$ for $^{18}$Ne,
the beam is so collimated that the event rates increase almost
linearly with increasing detector length in contrast to the $^8$B
case we have presented. However even in such an extreme scenario,
where we use the same storage ring configuration to reach such a
high $\gamma$, the expected event rates are comparable to that in
the $^8$B case. Hence we conclude that $^8$B is preferred to
$^{18}$Ne in exploring lepton number violating signatures with
beta-beams.\\

%%%%%%%%%%%%%%%%%% FIGURE 9,  EVENTS WITH 18_NEON FLUX %%%%%%%%%%%%%%%%%
%%%%%%%%%%%%%%%%%%%%%%%%%%%%%%%%%%%%%%%%%%%%%%%%%%%%%%%%%%%%%%%%%%%%%

The use of water \u{C}erenkov detectors with good capability of
muon-electron separation and moderate efficiency of neutral current
rejection may be an interesting option to see the signals of new
physics and to normalize the incoming flux. The disadvantage of this
set-up turns out to be the huge background.  Consider a 5 kT water
\u{C}erenkov detector with radius 2.5 m at a distance 200 m from the
decay ring. In five years, this will lead to 45 (12) muon events from
$\tau$-lepton decay for $\lp\lp$ ($\lambda\lp$) driven processes from
an incoming $^8$B $\nu_{e}$ beam accelerated with a $\gamma$ of 250
and with a muon detection threshold of 200 MeV. For the same
configuration and duration, one expects roughly $10^8$ pions produced
from charged and neutral current interactions of the $\nu_e$
beam. Muons produced from $\pi$ decay will thus completely swamp the
signal.\\

The number of signal events may be increased by designing a very long
water detector with small radius though this could be technologically
challenging.  In any case, the background events will continue to be
very high.  So, this option also does not hold much promise. The basic
problem of high backgrounds, avoided in the Fe detector, will also
plague totally active scintillator based detectors.

\section{Discussion and conclusion}

Beta-beam experiments may be sensitive to the lepton number violating
interactions. In~\cite{agarwalla2} it was shown that RPV $L$-violating
interactions can interfere with pure oscillation signals in
long-baseline beta-beam experiments. In this paper we explore a
complementary scenario. We propose that to probe such interactions, an
iron calorimeter detector placed close to the storage ring holds
promise as it provides essentially a neutrino oscillation free
environment. In particular, the combination of a 5 kT cylindrical iron
detector placed within a distance of 200 m to 1 km from the decay ring
and a neutrino beam from an $^{8}$B ion source with $\gamma$ in the
range 250 to 450, running for 5 years is well-suited in this
regard. We have examined the impact of non-trivial design details of
such a near-detector setup.\\

At production, low energy $\beta$-decay experiments may get
contaminated by tau neutrinos through RPV interactions.  We show that,
this contamination, though small, can be probed using the above
setup. RPV interactions can also play a role in such an experiment
during the interactions of the beta-beam electron neutrinos with the
detector.\\

It is interesting to explore if RPV interactions can affect beta-beam
experiments in other ways. For example, we have checked that the
impact of these interactions on the $\mu$ detection cross section is
insignificant.  As mentioned earlier, $\nu_\mu$ may be produced in
beta decay through RPV interactions but this also is severely
suppressed as the corresponding couplings have stringent upper
limits.\\

We have presented results for a neutrino beam. Anti-neutrino beams can
also be produced using $^8$Li or $^6$He as sources.  In fact, a
storage ring design may allow both beams to be present
simultaneously. The expected event rates for anti-neutrinos are of
similar order as for the neutrinos.\\

In conclusion, we find a near-detector setup can be useful for
exploring lepton number violating interactions with beta-beams.  It
may allow us to put stringent bounds on some of these couplings.\\

{\large{\bf {Acknowledgments}}}\\

SKA is grateful to N.K. Mondal for illuminating discussions. He would
like to acknowledge help from S.M. Shalgar and A. Samanta for
the GEANT and Nuance based  simulation. SR benefitted from
discussions with G. Majumder on pion background issues. He would like
to thank the Saha Institute of Nuclear Physics, Kolkata, India for a
fellowship in the early stages of this work.  He also acknowledges
support from `Bundesministerium f\"ur Bildung und Forschung',
Berlin/Bonn. SKA and AR acknowledge support from a DST, India research
project at the University of Calcutta during the initial period.

%%%%%%%%%%%%%%%%%%%%%%%%%%%%%%%%%%%%%%%%%%%%%%%%%%%%%%%%%%%%%%%%%%%%%%%%

%%%%%%%%%%%%%%%%%%%%%%%%%%%%%%%%%%%%%%%%%%%%%%%%%%%%%%%%%%%%%%%%%%%%%%%%
\end{document}